\documentclass[aps,prd,twocolumn,groupedaddress,superscriptaddress,preprintnumbers,nofootinbib]{revtex4}
\usepackage{graphicx}
\usepackage{epsfig}
\usepackage{rotating}
\usepackage{amssymb}
\usepackage{subfigure}
\usepackage{dsfont}
\usepackage{psfrag}
\usepackage{amsmath,euscript,array,mathrsfs}

\newcommand{\beq}{\begin{equation}}
\newcommand{\eeq}{\end{equation}}
\newcommand{\beqs}{\begin{eqnarray}}
\newcommand{\eeqs}{\end{eqnarray}}

\def\hbar{\hspace{0pt}\raisebox{1pt}{$-$} \hspace{-7pt} h}

\newcommand{\be}{\begin{equation}}
\newcommand{\ee}{\end{equation}}

\newcommand{\bea}{\begin{eqnarray}}
\newcommand{\eea}{\end{eqnarray}}

\def\lbldef#1#2{\expandafter\gdef\csname #1\endcsname {#2}}

\def\href#1#2{#2}

\newcommand{\ber}{\begin{eqnarray}}
\newcommand{\eer}{\end{eqnarray}}

\newcommand{\beqar}{\begin{eqnarray}}

\newcommand{\eeqar}{\end{eqnarray}}

\newcommand{\dsl}
  {\kern.06em\hbox{\raise.15ex\hbox{$/$}\kern-.56em\hbox{$\partial$}}}

\newcommand{\eeqarr}{\end{eqnarray}}
\newcommand{\ZZ}{{\rm \kern 0.275em Z \kern -0.92em Z}\;}

\def\CC{{\mathchoice
{\rm C\mkern-8mu\vrule height1.45ex depth-.05ex
width.05em\mkern9mu\kern-.05em}
{\rm C\mkern-8mu\vrule height1.45ex depth-.05ex
width.05em\mkern9mu\kern-.05em}
{\rm C\mkern-8mu\vrule height1ex depth-.07ex
width.035em\mkern9mu\kern-.035em}
{\rm C\mkern-8mu\vrule height.65ex depth-.1ex
width.025em\mkern8mu\kern-.025em}}}

\def\RR{{\rm I\kern-1.6pt {\rm R}}}

\def\ZZ{{\rm Z}\kern-3.8pt {\rm Z} \kern2pt}
\def\IB{\relax{\rm I\kern-.18em B}}
\def\ID{\relax{\rm I\kern-.18em D}}
\def\II{\relax{\rm I\kern-.18em I}}
\def\IP{\relax{\rm I\kern-.18em P}}

\newcommand{\bear}{\begin{eqnarray}}
\newcommand{\eear}{\end{eqnarray}}

\def\to{\rightarrow}

\def\to{\rightarrow}

\def\6{\partial}

\def\bea{\begin{eqnarray}}
\def\eea{\end{eqnarray}}

\def\beqx{\begin{displaymath}}
\def\eeqx{\end{displaymath}}

\newcommand{\bmat}{\left(\begin{array}}
\newcommand{\emat}{\end{array}\right)}

\def\bo{{\raise-.3ex\hbox{\large$\Box$}}}               
\def\face{{\raise.2ex\hbox{$\displaystyle \bigodot$}\mskip-2.2mu \llap {$\ddot
        \smile$}}}                                   
\def\>{\rangle}                                      
\def\<{\langle}                                      

\def\leftrightarrowfill{$\mathsurround=0pt \mathord\leftarrow \mkern-6mu
        \cleaders\hbox{$\mkern-2mu \mathord- \mkern-2mu$}\hfill
        \mkern-6mu \mathord\rightarrow$}        
\def\dvec#1{\vbox{\ialign{##\crcr
        \leftrightarrowfill\crcr\noalign{\kern-1pt\nointerlineskip}
        $\hfil\displaystyle{#1}\hfil$\crcr}}}           

\def\-{\hphantom{-}}

\begin{document}
\preprint{PNUTP-20/A02}
\title{Color dependence of tensor and scalar glueball masses in Yang-Mills theories}

\author{Ed Bennett}
\affiliation{Swansea Academy of Advanced Computing, Swansea University,
Bay Campus, SA1 8EN, Swansea, Wales, UK}
\author{Jack Holligan}
\affiliation{Department of Physics, College of Science, Swansea University,
Singleton Park, SA2 8PP, Swansea, Wales, UK}
\affiliation{The Institute for Computational Cosmology (ICC), Department of Physics, South Road, Durham, DH1 3LE, UK}
\author{Deog Ki Hong}
\affiliation{Department of Physics, Pusan National University, Busan 46241, Korea}
\author{Jong-Wan Lee}
\affiliation{Department of Physics, Pusan National University, Busan 46241, Korea}
\affiliation{Extreme Physics Institute, Pusan National University, Busan 46241, Korea}
\author{C.-J.~David~Lin}
\affiliation{Institute of Physics, National Chiao-Tung University, 1001 Ta-Hsueh Road, Hsinchu 30010, Taiwan}
\affiliation{Centre for High Energy Physics, Chung-Yuan Christian University,
Chung-Li 32023, Taiwan}
\author{Biagio Lucini}
\affiliation{Department of Mathematics, College of Science, Swansea University,
Bay Campus, SA1 8EN, Swansea, Wales, UK}
\affiliation{Swansea Academy of Advanced Computing, Swansea University,
Bay Campus, SA1 8EN, Swansea, Wales, UK}

\author{Maurizio Piai}
\affiliation{Department of Physics, College of Science, Swansea University,
Singleton Park, SA2 8PP, Swansea, Wales, UK}
\author{Davide Vadacchino}
\affiliation{INFN, Sezione di Pisa, Largo Pontecorvo 3, 56127 Pisa, Italy}


\date{\today}


\begin{abstract}

We report  the masses of the lightest spin-0 and spin-2 glueballs obtained in
  an extensive lattice study of the continuum and infinite volume limits of 
 $Sp(N_c)$ gauge theories for $N_c=2,4,6,8$. We also extrapolate the combined results towards
the large-$N_c$ limit.
We compute the ratio of scalar and tensor masses, and
observe evidence that this ratio is independent of $N_{c}$.  
Other lattice studies  of Yang-Mills theories at the same space-time dimension
provide a compatible ratio.
We further compare these results to various analytical ones 
and discuss them in view of symmetry-based arguments related to the breaking of scale invariance 
in the underlying dynamics, showing that a constant ratio might emerge 
in a scenario in which the $0^{++}$ glueball is interpreted as a dilaton state. 

\end{abstract}


\maketitle
\section{Introduction}

In $D=3+1$ space-time dimensions, Yang-Mills (YM) theories 
are classically scale-invariant. At high energies the theory is perturbative, and
governed by a trivial fixed point---this is the essence of asymptotic freedom.
Scale symmetry is anomalous though, broken by quantum effects that make the theory
flow away from its trivial fixed point, and introduce an intrinsic scale $\Lambda$, via dimensional transmutation.

At high energy, the massless gluons, carrying color charges, are the natural choice of degrees of freedom
to describe small perturbations around the trivial fixed point.
Yang-Mills theories are believed to confine at low energies ${\cal O}(\Lambda)$.
Low-energy excitations are color singlets, called glueballs, 
and their spectrum is  gapped.
 The phenomena associated with the transition to the confined phase are  intrinsically 
 non-perturbative and   difficult to study.

In Ref.~\cite{Bennett:2017kga}, some of us started an extensive 
study of $Sp(N_c)$ gauge theories, which includes calculating the masses of
the glueballs in the YM theory.
The spectrum of $Sp(4)$ glueballs was one of the most robust results of that exploratory and 
agenda setting paper. 
We update the measurements for the $Sp(4)$ group, by doubling the size of the combined statistical ensemble, 
and then proceed to the next step of this programme, by performing detailed studies 
of the YM theory (with no matter content) with gauge groups $Sp(2)$, $Sp(6)$, and $Sp(8)$ (see also preliminary results 
in Ref.~\cite{Holligan:2019lma}).
We report here our results for the
lightest scalar and tensor glueballs.

Understanding  the glueball spectrum is tantamount to solving the  YM theory, 
and  uncovering  the mechanism of confinement.
Reference~\cite{Athenodorou:2016ndx}  suggested that the quantity
\beqs
R&\equiv&\frac{m_{2^{++}}}{m_{0^{++}}}\,,
\label{Eq:R}
\eeqs
defined as the ratio of masses of the glueballs with quantum number $J^{PC}=2^{++}$ and $J^{PC}=0^{++}$,
captures some  universal, intrinsic properties of YM theories, in the sense that it 
depends only on the dimensionality of the space-time and 
of the operators of the field theory.
We devote this paper to these specific observables.
A comprehensive report on the 
physics of $Sp(N_c)$ YM theories, which details the results for excited states and for 
extended objects, is in preparation~\cite{us}.


\section{Glueball masses: new lattice results}

We report at the top of Table~\ref{Fig:data}  our new lattice measurements of 
glueball masses in $D=3+1$ dimensions for  $Sp(N_c)$ YM theories.
 The algorithm employed in our lattice calculations adopts the Wilson action,
 and the local updates are based upon a combination of Heat Bath and Over Relaxation, 
by supplementing  the Cabibbo-Marinari update with a simple re-symplectisation procedure,
as described in Ref.~\cite{Bennett:2017kga}.

\begin{widetext}

\begin{table}[ht]
\begin{center}
\begin{tabular}{|c|c|c|c|c|c|c|c|}
\hline\hline
 &&&&&&& \cr
$D$ & {\rm ~~~Group~~~} & Reference  &  $\frac{m_{0^{++}}}{\sqrt{\sigma}}=\frac{m_{A_1^{++}}}{\sqrt{\sigma}} $ &  $\frac{m_{E^{++}}}{\sqrt{\sigma}}$ &  
$\frac{m_{T_2^{++}}}{\sqrt{\sigma}}$ & $\frac{m_{2^{++}}}{\sqrt{\sigma}}$ & $R$ \cr
 &&&&&&& \cr
\hline
$3+1$ & $Sp(2)$ & \cite{us}                                 & $\bf{3.841(84)}   $ & ${\bf 5.33(18)}   $  & ${\bf 5.29(20)}   $  & ${\bf  5.31(13)}   $  & ${\bf 1.383(46)}$ \cr
$3+1$ & $Sp(4)$ & \cite{Bennett:2017kga,us} & $ \bf {3.729(89)}$&$\bf{  5.14(16)}$&  $\bf{5.03(18)}$ &  $\bf{5.09(12)}$ & $\bf{1.366(45)}$ \cr
$3+1$ & $Sp(6)$ & \cite{us}                                 & $\bf{3.430(75)}   $ & ${\bf 5.03(13)}   $  & ${\bf 5.09(16)}    $ & ${\bf  5.05(10)}   $  & ${\bf 1.473(43)}$ \cr
$3+1$ & $Sp(8)$ & \cite{us}                                 & $\bf{3.308(98)}   $ & ${\bf 4.62(29)}   $  & ${\bf 4.73(23)}    $ &${\bf  4.69(18)}   $   &${\bf 1.417(69)}$  \cr
$3+1$ & $Sp(\infty)$ & \cite{us}                      &$\bf{3.241(88)}$&$\bf{ 4.79(19) }$&$\bf{ 4.80(20) }$&$\bf{ 4.80(14)}$&$\bf{  1.480(58)}$
\cr
\hline
$3+1$ & $SU(2)$ & Table 14~\cite{Lucini:2004my} & $3.78(7)   $ & - & -  & $5.45(11)  $ & ${\bf 1.442(39)}$\cr
$3+1$ & $SU(3)$ & Table 14~\cite{Lucini:2004my} & $3.55(7)   $ & -  & -  & $4.78(9)  $ & ${\bf 1.346(37)}$\cr
$3+1$ & $SU(4)$ & Table 14~\cite{Lucini:2004my} & $3.36(6)   $ & -  & -  & $4.88(11)  $ & ${\bf 1.452(42)}$\cr
$3+1$ & $SU(6)$ & Table 14~\cite{Lucini:2004my} & $3.25(9)   $ & -  & -  & $4.73(15)  $ &${\bf 1.455(61)}$ \cr
$3+1$ & $SU(8)$ & Table 14~\cite{Lucini:2004my} & $3.55(12)   $ & -  &-  & $4.73(22)  $ & ${\bf 1.332(77)}$\cr
$3+1$ & $SU(\infty)$ & Table 14~\cite{Lucini:2004my} & $3.307(53)   $ & -  & -  & $4.80(14)  $ & ${\bf 1.451(48)}$\cr
\hline
$2+1$ & $SO(3)$ & Table 28~\cite{Lau:2017aom} & $3.132(34)   $ & - & -  & $5.13(9)  $ & ${\bf 1.638(34)}$\cr
$2+1$ & $SO(4)$ & Table 28~\cite{Lau:2017aom} & $3.343(23)   $ & - & -  & $5.711(81)  $ & ${\bf 1.708(27)}$\cr
$2+1$ & $SO(5)$ & Table 28~\cite{Lau:2017aom} & $3.545(17)   $ & - & -  & $6.008(46)  $ & ${\bf 1.695(15)}$\cr
$2+1$ & $SO(6)$ & Table 28~\cite{Lau:2017aom} & $3.656(13)   $ & - & -  & $6.190(38)  $ & ${\bf 1.693(12)}$\cr
$2+1$ & $SO(7)$ & Table 29~\cite{Lau:2017aom} & $3.737(10)   $ & - & -  & $6.297(54)  $ & ${\bf 1.685(15)}$\cr
$2+1$ & $SO(8)$ & Table 29~\cite{Lau:2017aom} & $3.788(14)   $ & - & -  & $6.498(36)  $ & ${\bf 1.715(11)}$\cr
$2+1$ & $SO(12)$ & Table 29~\cite{Lau:2017aom} & $3.878(24)   $ & - & -  & $6.636(64)  $ & ${\bf 1.711(20)}$\cr
$2+1$ & $SO(16)$ & Table 29~\cite{Lau:2017aom} & $3.973(15)   $ & - & -  & $6.714(40)  $ & ${\bf 1.690(12)}$\cr
$2+1$ & $SO(\infty)$ & Table 31~\cite{Lau:2017aom} & $4.150(33)   $ & - & -  & $6.987(88)  $ & ${\bf 1.684(25)}$\cr
$2+1$ & $SO(\infty)$ & Table 31~\cite{Lau:2017aom} & $4.179(16)   $ & - & -  & $7.129(43)  $ & ${\bf 1.706(12)}$\cr
\hline
$2+1$ & $SU(2)$ & Table B3~\cite{Athenodorou:2016ebg}  & $4.7369(55)   $ & - & -  & $7.762(10)  $ & ${\bf 1.6386(28)}$\cr
$2+1$ & $SU(3)$ & Table B4~\cite{Athenodorou:2016ebg}  & $4.3683(73)   $ & - & -  & $7.241(17)  $ & ${\bf 1.6576(48)}$\cr
$2+1$ & $SU(4)$ & Table B5~\cite{Athenodorou:2016ebg} & $4.242(9)   $ & - & -  & $7.091(17)  $ & ${\bf 1.6616(54)}$\cr
$2+1$ & $SU(6)$ & Table B6~\cite{Athenodorou:2016ebg}  & $4.164(8)   $ & - & -  & $6.983(19)  $ & ${\bf 1.6770(56)}$\cr
$2+1$ & $SU(8)$ & Table B7~\cite{Athenodorou:2016ebg} & $4.144(10)   $ & - & -  & $6.952(18)  $ & ${\bf 1.6776(59)}$\cr
$2+1$ & $SU(12)$ & Table B8~\cite{Athenodorou:2016ebg} & $4.140(9)   $ & - & -  & $6.938(18)  $ & ${\bf 1.6759(57)}$\cr
$2+1$ & $SU(16)$ & Table B9~\cite{Athenodorou:2016ebg}  & $4.129(11)   $ & - & -  & $6.937(30)  $ & ${\bf 1.6801(85)}$\cr
$2+1$ & $SU(\infty)$ & Tables B10,B11~\cite{Athenodorou:2016ebg} & $4.116(6)   $ & - & -  & $6.914(13)  $ & ${\bf 1.6798(40)}$\cr
\hline\hline
\end{tabular}
\end{center}
\caption{Lattice measurements of the masses of the glueballs, as described in the main text.
 In bold face are the calculations performed for this letter, while the other numerical values are lifted from the literature, as indicated.
 In the case of $Sp(4)$, new measurements have been combined with those from  Ref.~\cite{Bennett:2017kga}, doubling the combined statistics.
}
\label{Fig:data}
\end{table}

\end{widetext}

We restrict attention to the ratio $m_G/\sqrt{\sigma}$ between glueball masses $m_G$ and the 
square root of the string tension $\sigma$. 
The notation $G={E^{++}, A_1^{++}, T_2^{++}}$, refers explicitly to the representations
of the octahedral group, which describes the symmetry of the discretised space-time, and to $P$ and $C$ quantum numbers,
as in Ref.~\cite{Berg:1982kp}. 
In the  measurements,  we combine  the smearing and blocking of Ref.~\cite{Lucini:2004my}
with the extended basis of operators in the variational approach of Ref.~\cite{Lucini:2010nv}.

 The errors  are due 
 to statistical uncertainties.  We perform continuum-limit extrapolations with a
conventional linear fit to the dependence on $a^2$, where $a$ is the lattice spacing.
We  also report a simple large-$N_c$ extrapolation, in which we include corrections ${\cal O}(1/N_c)$  to $m_G/\sqrt{\sigma}$, since the leading corrections occurs at $1/N_c$~\cite{Lovelace:1982hz}.
 We find that the uncertainty in
 the string tension $\sigma$ is much smaller than in the masses $m_G$.
 Other technical details, including 
comments on the systematics and on finite size effects,
 will appear in Ref.~\cite{us}.

We identify $m_{A_1^{++}}=m_{0^{++}}$. As $m_{E^{++}}$ and $m_{T_2^{++}}$ are compatible with each other,
and they both relate to the symmetric tensors in the continuum theory~\cite{Lucini:2010nv},
we compute $m_{2^{++}}$ as the weighted average of the two. Finally,
the error on the ratio $R$ is obtained by simple propagation. The error is overestimated, as we ignore correlations, in particular because
of the common dependence on $\sigma$, but we expect such effects to be small, and not to affect our discussion.

Figure~\ref{Fig:R} shows that
the ratio $R$ for the sequence of $Sp(N_c)$ YM theories  is compatible with a constant.
This confirms that $O(1/N_c)$
effects, if present, are smaller than the current  uncertainties,
the magnitude of which varies between $\sim 2\%$ for $Sp(4)$  and $5\%$ for $Sp(8)$.


\section{Glueball masses: earlier lattice results}

\begin{figure}[ht]
\begin{center}
\begin{picture}(240,210)
\includegraphics[height=6.9cm]{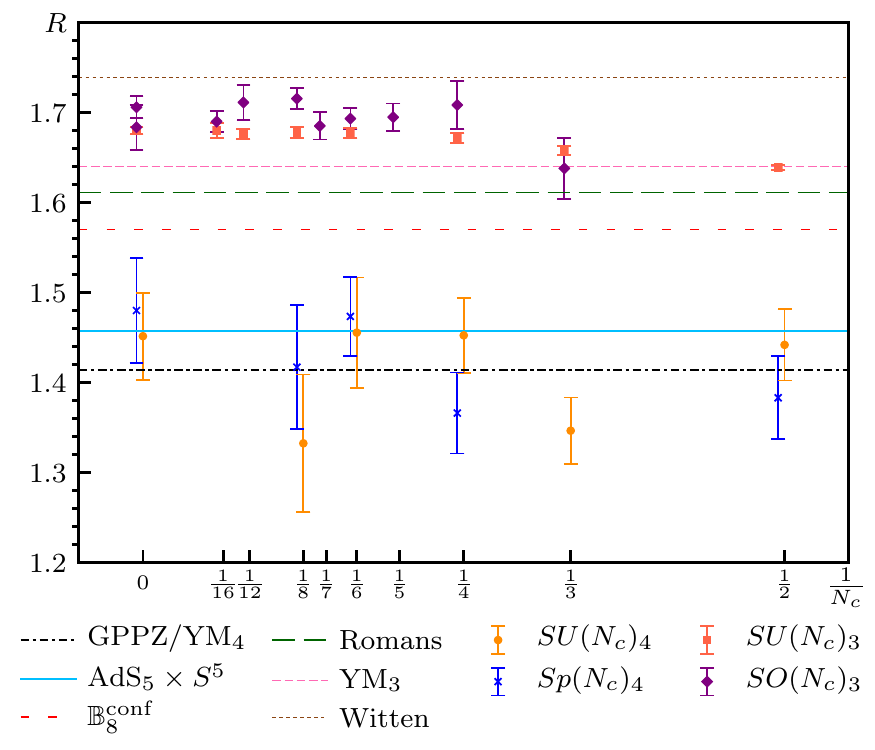}
\end{picture}
\caption{Numerical and analytical results for the ratio $R$ defined in Eq.~(\ref{Eq:R}). Different
 shaped markers denote the lattice measurements with continuum extrapolations in $D=3+1$ dimensions for $Sp(N_c)$  
 and for $SU(N_c)$~\cite{Lucini:2004my}, as well as in $D=2+1$ dimensions for 
 $SO(N_c)$~\cite{Lau:2017aom} and $SU(N_c)$~\cite{Athenodorou:2016ebg}. 
 Extrapolations to the $N_c\rightarrow \infty$ limit are also included. 
 Differently rendered lines at $R=\sqrt{2}, 1.46, 1.57,  1.61,  1.74$, are the holographic 
 calculations in the GPPZ model~\cite{Mueck:2004qg}, the circle reduction of AdS$_5\times S^5$~\cite{Brower:2000rp,Elander:2020csd}, 
 the holographic model $\mathbb{B}_8^{\rm conf}$ in Ref.~\cite{Elander:2018gte},
the Witten model~\cite{Brower:2000rp,Elander:2013jqa}, and
the circle reduction of Romans supergravity~\cite{Wen:2004qh,Elander:2013jqa}, respectively. 
 With $R=\sqrt{2}, 1.64$ we report the field theoretical results from 
 Refs.~\cite{Bochicchio:2013sra} and~\cite{Leigh:2006vg}, for YM theories in $D=3+1$ and $D=2+1$ dimensions, respectively. More details can be found in the main text.}
\label{Fig:R}
\end{center}
\end{figure}

We include in Table~\ref{Fig:data} and Figure~\ref{Fig:R} our measurements (denoted $Sp(N_c)_4$),
together with lattice results by other collaborations,
 for various classes of YM theories.

The spectrum of YM glueballs in $D=3+1$ dimensions with $SU(N_c)$ group (denoted $SU(N_c)_4$)
was studied in Refs.~\cite{Lucini:2010nv,Lucini:2004my}.
In the former, 
the authors use a single value of the lattice parameters for each  value of $N_c$,
without studying the approach to the continuum limit. 
Conversely, Ref.~\cite{Lucini:2004my} reports continuum limits for the glueball masses
expressed in units of the string tension $\sigma$,
but the variational method uses a smaller basis of operators of the octahedral group in respect to our work,
and the $T_2$ channel is not measured. 
As long as we restrict attention to the lightest states in the spectrum
 (the $0^{++}$ and $2^{++}$ ground states), at the same lattice spacing
the results of the two approaches
are in good agreement, and hence we compare the $Sp(N_c)$ sequence of
measurements, as well as their extrapolation to large $N_c$,
to those of Ref.~\cite{Lucini:2004my}.
As visible in Fig.~\ref{Fig:R}, the agreement in the ratio $R$ across the gauge groups is excellent.

We also summarise the lattice measurements for $SO(N_c)$ in $D=2+1$ dimensions ($SO(N_c)_3$), taken from 
 Tables~28, 29 and 31 of Ref.~\cite{Lau:2017aom} (see also Fig.~26 therein).
We include only continuum limit results, and two different types of large-$N_c$ extrapolations. 
Finally, we collect results for $SU(N_c)$ theories in $D=2+1$ dimensions  ($SU(N_c)_3$)
from Tables B3-B11 of Ref.~\cite{Athenodorou:2016ebg}. 
The extrapolation to $SU(\infty)$ has been performed by including  $1/N_c^2$ as well as $1/N_c^4$  corrections.

Lattice results on $R$ show the emergence of a regular pattern, that depends only on the dimensionality $D$ of the system.
The group sequence ($SU(N_c)$,  $Sp(N_c)$ or $SO(N_c)$) and the number of colors $N_c$ do not appear to affect $R$, 
within current uncertainties---with some deviation from this pattern in $D=2+1$ dimensions for $SU(3)$, $SO(3)$ and $SU(2)$.
We have at our disposal preliminary results for excited states and states with different quantum numbers
 in $Sp(N_c)$ theories (to appear in Ref.~\cite{us}), and we did not find significant evidence of similar regular patterns,
reinforcing the notion that the lightest $0^{++}$ and $2^{++}$ 
glueballs play a special role in YM theories.

\section{Glueball masses: a brief survey of analytical results}

In Fig.~\ref{Fig:R}, we compare the result  of lattice measurements of the ratio $R$ to 
two classes of semi-analytical calculations,
performed either via gauge-gravity dualities arising in the context of supergravity,
or via alternative field-theory methods.
In all these models, the ratio $R$ is known only in the strict large-$N_c$ limit, as
$1/N_c$ corrections are ignored.

The GPPZ model was proposed in Ref.~\cite{Girardello:1999bd}
 (see also Refs.~\cite{Girardello:1998pd,Distler:1998gb,Pilch:2000fu}) as a simple, 
 classical supergravity dual of 
mass-deformed, large-$N_c$, ${\cal N}=4$ Super-Yang-Mills. The geometry is singular and  asymptotically 
approaches AdS$_5$.
The spectrum of fluctuations 
 yields $R=\sqrt{2}$~\cite{Mueck:2004qg} (see also Refs.~\cite{Apreda:2003sy,Elander:2011aa,Elander:2012fk}). 
This result happens to be in exact
agreement  with that of the large-$N_c$ field-theory study in Ref.~\cite{Bochicchio:2013sra} (see Table~1 therein),
which in  Fig.~\ref{Fig:R} we denote as YM$_4$.
A closely related model is studied in Ref.~\cite{Brower:2000rp}, that reports a  holographic  calculation
based upon the circle reduction of the system yielding the 
AdS$_5\times S^5$ background (see also Ref.~\cite{Elander:2020csd}).
The result in this case is $R=1.46$.
The close proximity between the results of these two holographic calculations (both of which use
geometries that are asymptotically AdS$_5$),  Bochicchio's field-theoretical 
approach~\cite{Bochicchio:2013sra,Bochicchio:2016toi}, and lattice calculations in $Sp(N_c)$ and $SU(N_c)$ is remarkable.

 Witten's holographic model of confinement~\cite{Witten:1998zw}
is based upon $S^1\times S^1\times S^4$ reduction of  eleven-dimensional 
supergravity~\cite{Nastase:1999cb,Pernici:1984xx,Pernici:1984zw,Lu:1999bc}.
In the asymptotically AdS$_7$ background geometry,
 one $S^1$ shrinks to zero size. 
   The static quark-antiquark potential
is computed holographically~\cite{Maldacena:1998im,Rey:1998ik},
and yields linear confinement.
Adaptations  to model quenched QCD were proposed in Refs.~\cite{Sakai:2004cn,Sakai:2005yt}.
The spectrum of glueballs yields $R=1.74$~\cite{Brower:2000rp} (see also Ref.~\cite{Elander:2013jqa}).  
An alternative model, based on circle reduction of Romans supergravity~\cite{Romans:1985tw}, has geometry that 
is asymptotically AdS$_6$, and again the circle shrinks. 
In this case, $R=1.61$~\cite{Wen:2004qh} (see also Refs.~\cite{Kuperstein:2004yf,Elander:2013jqa,Elander:2018aub}).
For both celebrated models, Fig.~\ref{Fig:R} shows that $R$ is not compatible with the lattice results, with current uncertainties.

The literature on the holographic dual of three-dimensional confining theories is more limited.
In Ref.~\cite{Elander:2018gte}
the model dubbed $\mathbb{B}_8^{\rm conf}$
is the gravity dual of a non-trivial, asymptotically free theory in $2+1$ dimensions~\cite{Gibbons:1989er,Hashimoto:2010bq,Cvetic:2001ye,Faedo:2017fbv},
and yields $R\simeq 1.57$.
A completely different field-theory approach to
YM theories in $2+1$ dimensions is used to compute glueball masses
in Refs.~\cite{Leigh:2005dg,Leigh:2006vg} (we denote it as YM$_3$ in Fig.~\ref{Fig:R}). 
From the latter of the two, we read that $R\simeq 1.64$.
This result is valid only in the strict $N_c\rightarrow +\infty$ limit, although the analysis in Ref.~\cite{Leigh:2006vg}
could potentially be extended to finite $N_c$.
Both these approaches ($\mathbb{B}_8^{\rm conf}$ and YM$_3$ in Fig.~\ref{Fig:R}) slightly underestimate $R$
in respect to the lattice results for $SU(N_c)$ and $SO(N_c)$.

\section{Discussions and Universal Ratio}
\label{Sec:argument}

If the  ratio between  the masses of the lightest spin-2 and spin-0 glueballs is universal for (pure) 
YM theories, there should be underlying principles that hold for all of them. 
We argue (see also Ref.~\cite{Hong:2017suj}) that scale symmetry and perturbative unitarity are such principles.

When the YM theory undergoes the phase transition to the confining phase, the vacuum energy density  ${\cal E}_{\rm vac}$
 is lowered, breaking  scale invariance spontaneously, to yield
 \begin{equation}
 {\cal E}_{\rm vac}\equiv \frac14\left<T^{\mu}_{\mu}\right><0\,,
 \end{equation}
 with $T_{\mu\nu}$ the energy-momentum tensor.

As the vacuum is not invariant under  scale transformations, 
the dilatation current $D_{\mu}=x^{\nu}T_{\mu\nu}$ creates a state, called a dilaton,  out of the vacuum,
which we write as 
\begin{equation}
\left<0\right|D_{\mu}(x)\left|\sigma(p)\right>\equiv if_Dp_{\mu}e^{-ip\cdot x}\, ,
\end{equation}
where $f_D$ is the dilaton decay constant. 
 If the two-point function of dilatation currents 
 is dominated by the dilaton pole at low energy,  for $p\to0$ we expect:
\begin{equation}
\int_x\,e^{ip\cdot x}\langle 0 |{\rm T} [T^{\mu}_{\mu}(x)T^{\nu}_{\nu}(0)] |0\rangle\approx f_D^2m_D^2\,=-16\, {\cal E}_{\rm vac}\,,
\label{pcdc}
\end{equation}
with $m_D$ being the dilaton mass.
Under this assumption, we identify the ground-state glueball with the dilaton, 
because it is the lightest particle and both of them have the same quantum numbers as the vacuum.
How good this approximation is can only be assessed a posteriori.

The  Lagrangian density of the dilaton low-energy effective field theory (EFT) is the subject of a vast literature.  The potential
must break scale invariance explicitly, and contain non-marginal operators.
Departures from marginality might be encoded
in a logarithmic field-dependent potential, as advocated  in Refs.~\cite{Schechter:1980ak,Migdal:1982jp}.
(More general, power-law potentials have also been considered~\cite{Rattazzi:2000hs,Chacko:2012sy,
Appelquist:2017wcg,Appelquist:2017vyy,Cata:2018wzl,Cata:2019edh,Appelquist:2019lgk,Fodor:2020niv}).
We dispense with such level of detail in the context of this discussion. 
It is natural to assume that the intrinsic, dynamically generated scale $\Lambda$
sets ${\cal E}_{\rm vac}\sim \Lambda^4$ and $f_D\sim\Lambda$. 
Therefore, from Eq.~(\ref{pcdc}) and taking $16{\cal E}_{\rm vac}=-\beta f_D^4$,
we may write
\begin{equation}
f_D^2m_D^2=\beta f_D^4\,.
\label{dilaton_mass}
\end{equation}
The numerical constant $\beta$ is an intrinsic constant of the YM theory, and depends on the gauge group.
It measures the size of explicit breaking of scale symmetry, sets the strength of the self-interaction of the dilaton,
 and is the expansion parameter of the EFT. The  parameter $\beta$ is not 
 guaranteed to be  small.
Lattice calculations find that  the spin-2 glueball is the lowest excited state, and has mass 
of the same order of magnitude as that of the ground-state glueball.  

The dilaton EFT  yields the amplitude ${\cal M}_{\sigma}$, for the scattering process
 $\sigma(p_1)+\sigma(p_2)\to \sigma(p_3)+\sigma(p_4)$ between dilaton particles. 
 For  center-of-mass energies $E\gg m_D$, 
 we borrow  Eq.~(3.3) from Ref.~\cite{Komargodski:2011vj} (see also Ref.~\cite{dkh})
and write
\begin{equation}
{\cal M}_{\sigma}\sim-\frac{1}{\alpha^4f_D^4}\left( s^2+t^2+u^2\right)+{\cal O}\left(\frac{m_D^2}{f_D^2}\right)\,,
\end{equation} 
in terms of the Mandelstam variables  $s=(p_1+p_2)^2$, $t=(p_3-p_1)^2$, and $u=(p_4-p_1)^2$\,.  
Here $\alpha$ is a dimensionless constant characterising the theory.
The scattering amplitude violates perturbative unitarity at $E\sim \alpha f_D$,  
To achieve partial unitarity restoration, and raise this bound, we introduce the spin-2 glueball  in the EFT. 
We assume that the spin-2 glueball couples to the energy-momentum tensor of the dilaton
$T_D^{\mu\nu}$.

The Lagrangian density of the massive spin-2 glueball $h_{\mu\nu}$ can be derived by
identifying it with the
expansion of the spacetime metric around the flat spacetime as in $g_{\mu\nu}=\eta_{\mu\nu}+2\kappa h_{\mu\nu}$, to obtain
\begin{equation}
{\cal L}_G={\cal L}_G^{\rm kin}-{\kappa}\,h_{\mu\nu}T^{\mu\nu}_D+\cdots\,,
\end{equation}
where the first term is the so-called Fierz-Pauli kinetic-term for the massive spin-2 fields, 
$\kappa$ is the (universal) coupling of the spin-2 glueballs and
 the ellipsis denotes the higher order terms. Again, the assumptions
 underneath this identification can be assessed a posteriori.
 
 The propagator of the massive spin-2 field of mass $m_T$  is then given by~\cite{Fierz:1939ix} 
\begin{equation}
\int_xe^{ip\cdot x}\left<0\right|T\left\{h_{\mu\nu}(x)h_{\alpha\beta}(0)\right\}\left|0\right>=\frac{iP_{\mu\nu\alpha\beta}}{p^2-m_T^2+i\epsilon}\,,
\end{equation}
where $2P_{\mu\nu\alpha\beta}=\tilde\eta_{\mu\alpha}
{\tilde \eta}_{\nu\beta}+{\tilde \eta}_{\mu\beta}{\tilde \eta}_{\nu\alpha}-\frac23{\tilde \eta}_{\mu\nu}{\tilde \eta}_{\alpha\beta}$ 
with
${\tilde \eta}_{\mu\nu}=\eta_{\mu\nu}-p_{\mu}p_{\nu}/m_T^2$.
The contribution of the diagrams with internal exchange of the spin-2 particles
changes the structure of the amplitude, and partially restores perturbative unitarity to hold 
at the scale $E\sim (\kappa f_D)^{-1}\cdot m_T $ and slightly above, where $\kappa f_D$ 
measures the strength of the spin-2 coupling to the dilaton, compared to the dilaton self-coupling.
For this to happen, one must require that $\alpha f_D\sim(\kappa f_D)^{-1}\cdot m_T $,
 or $m_T^2 \equiv g f_D^2\sim \alpha\kappa^2f_D^4$.

The dimensionless 
constant $g\sim\alpha\kappa^2f_D^2$ depends on the
 microscopic details of the theory,  as $\beta$.
Combining this with Eq.~(\ref{dilaton_mass}), 
we write the mass ratio of the spin-2 glueball and the ground-state glueball as
\begin{equation}
R^2\equiv\frac{m_T^2}{m_{D}^2}=\frac{g}{\beta}\,.
\end{equation}
In the mass ratio between the lightest 
 spin-2  and spin-0 glueball the dependence on microscopic details should decouple as suggested by the lattice data.
As the EFT captures the long-distance dynamics based on  symmetry (and 
perturbative unitarity) considerations, 
that are common to all YM theories,
 it should describe  all low-energy (pure) YM theories.

The lattice data we summarised suggests the ratio $R$ in $D = 2+1$ is also universal.  
It has been noted elsewhere that the similarities between 
the physics of confinement in $D = 2 + 1$ and in $D = 3 + 1$ dimensions
turn out to be much deeper than naively expected (see e.g. Ref.~\cite{Teper:1998te}). 
On this basis, we argue that also in $D = 2 + 1$ dimensions
the constant ratio is controlled by spontaneous as well as explicit 
breaking of scale invariance through confinement, 
which, by  generating  a mass gap, changes the would-be 
power law behaviour of gluon correlators, 
at distances much larger than the intrinsic length scale set by the 
dimensional gauge coupling.

\section{Outlook}

Our 
lattice measurements of the masses of the lightest scalar and tensor glueballs for $Sp(N_c)$ gauge theories
in $D=3+1$ dimensions show no discernible dependence on $N_c$ in the ratio $R$ defined by Eq.~(\ref{Eq:R}).
We compared this finding with lattice measurements taken from the literature,
and compiled a (non exhaustive) list of other calculations, that use holography or alternative 
field theory methods.
We found supporting empirical evidence that the ratio  $R$ might be a universal quantity in YM theories, 
in the sense that it appears to depend only on the dimensionality of the system, not its microscopic details.

This intriguing feature might be connected with the special role that the lightest scalar glueball
and the lightest tensor glueball play in respect to scale invariance. 
As we argued in Section~\ref{Sec:argument}, it might be explained 
under the approximation that these two particles can be identified with those
sourced by the dilatation operator and by the energy-momentum tensor.
This approximation relies on two separate assumptions: that the explicit breaking 
of scale invariance is small compared to its spontaneous breaking, 
and that  single particle exchange saturates the 2-point correlation functions build
with the dilatation operator and the energy-momentum tensor.

Our arguments  highlight the distinguishing features of the two
particles that are the main topic of this letter.
More theoretical work would be useful, to better understand the role of these two particles, and
whether the empirical evidence we uncovered points to an exact relation, 
or, if otherwise, to estimate the size of deviations.
It would also be very useful to have lattice data on Yang-Mills theories with other gauge groups,
and we hope  such calculations will be performed in the future.

\vspace{0.0cm}
\begin{acknowledgments}
\end{acknowledgments}

We thank D.~Elander for discussion about Ref.~\cite{Elander:2018gte}.

The work of EB has been funded  by the Supercomputing Wales project, 
which is part-funded by the European Regional Development Fund (ERDF) via Welsh Government. 

JH is supported by the STFC Consolidated Grant ST/P00055X/1, by the College of Science, Swansea University,
and  by the STFC-DTG  ST/R505158/1.

The work of DKH was supported by Basic Science Research Program 
through the National Research Foundation of Korea (NRF) funded by 
the Ministry of Education (NRF-2017R1D1A1B06033701) and he is grateful to CERN-TH for the hospitality, where the part of this work is done.  

The work of JWL is supported in part by the National Research Foundation of Korea grant funded 
by the Korea government(MSIT) (NRF-2018R1C1B3001379) and 
in part by Korea Research Fellowship programme funded 
by the Ministry of Science, ICT and Future Planning through the National 
Research Foundation of Korea (2016H1D3A1909283).

The work of CJDL is supported by the  Taiwanese MoST grant 105-2628-M-009-003-MY4. 

The work of BL and MP has been supported in part by the STFC 
Consolidated Grants ST/L000369/1 and ST/P00055X/1. BL and MP
received funding from the European Research Council (ERC) under the
European Union's Horizon 2020 research and innovation programme under
grant agreement No 813942.
The work of BL is further supported in part by the Royal Society Wolfson Research Merit Award WM170010.

DV is supported by the INFN HPC-HTC project.

Numerical simulations have been performed on the Swansea SUNBIRD 
system, on the local HPC
clusters in Pusan National University (PNU) and in National Chiao-Tung University (NCTU),
 and on the Cambridge Service for Data Driven Discovery (CSD3). The Swansea SUNBIRD 
system is part of the Supercomputing Wales project, which is part-funded by the European Regional
Development Fund (ERDF) via Welsh Government. CSD3 is operated in part by
 the University of Cambridge Research Computing on
behalf of the STFC DiRAC HPC Facility (www.dirac.ac.uk). 
The DiRAC component of CSD3 was funded by BEIS capital funding via
STFC capital grants ST/P002307/1 and ST/R002452/1 and 
STFC operations grant ST/R00689X/1. DiRAC is part of the National e-Infrastructure.



\end{document}